\documentclass[preprint,showpacs,preprintnumbers,amsmath,amssymb, showkeys,superscriptaddress]{revtex4}
\usepackage{graphicx}
\usepackage{dcolumn}
\usepackage{bm}
\usepackage{hyperref}

\begin{document}

\title{The Hoyle State in Relativistic ${}^{12}$C Dissociation \footnote[1]{Submitted to Proceedings of  XXII International Conference on Few-Body Problems in Physics (FB22), 9 - 13 July, 2018, Caen, France.}}

\author{D. A. Artemenkov}
\affiliation{Joint Institute for Nuclear Research (JINR), Dubna, Russia}
\author{M. Haiduc}
\affiliation{Institute of Space Science, Magurele, Romania}
\author{N.K. Kornegrutsa}
\affiliation{Joint Institute for Nuclear Research (JINR), Dubna, Russia}
\author{E. Mitsova}
\affiliation{Joint Institute for Nuclear Research (JINR), Dubna, Russia}
\affiliation{Southwestern University, Blagoevgrad, Bulgaria}
\author{N.G. Peresadko}
\affiliation{Lebedev Physical Institute, Russian Academy of science, Moscow, Russia}
\author{V.V. Rusakova}
\affiliation{Joint Institute for Nuclear Research (JINR), Dubna, Russia}
\author{R. Stanoeva}
\affiliation{Southwestern University, Blagoevgrad, Bulgaria}
\affiliation{Institute for Nuclear Research and Nuclear Energy, Sofia, Bulgaria}
\author{A.A. Zaitsev}
\affiliation{Joint Institute for Nuclear Research (JINR), Dubna, Russia}
\affiliation{Lebedev Physical Institute, Russian Academy of science, Moscow, Russia}
\author{P.I. Zarubin}
\email{zarubin@lhe.jinr.ru}
\affiliation{Joint Institute for Nuclear Research (JINR), Dubna, Russia}
\affiliation{Lebedev Physical Institute, Russian Academy of science, Moscow, Russia}
\author{I.G. Zarubina}
\affiliation{Joint Institute for Nuclear Research (JINR), Dubna, Russia}

\begin{abstract}
	Production of $\alpha$-particle triples in the Hoyle state (HS) in dissociation of ${}^{12}$C nuclei at 3.65 and 0.42 $A$ GeV in nuclear track emulsion is revealed by the invariant mass approach. Contribution of the HS to the dissociation ${}^{12}$C $\to$ 3$\alpha$ is (11 $\pm$ 3) \%. Reanalysis of data on coherent dissociation ${}^{16}$O $\to$ 4$\alpha$ at 3.65 $A$ GeV is revealed the HS contribution of (22 $\pm$ 2) \%. 
\end{abstract}

\maketitle

Events of dissociation of relativistic light nuclei observable in detail in the nuclear track emulsion (NTE) contain holistic information on ensembles of lightest nuclei which is of interest to the nuclear cluster physics \cite{1}. The best spatial resolution provided by the NTE technique turns out to be a decisive factor for recognition relativistic ${}^{8}$Be and ${}^{9}$B decays among the projectile fragments \cite{2}. The decays are identified by the invariant mass $M^*$ defined by the sum of all products of 4-momenta $P_i$ of relativistic fragments He and H. Subtracting the sum of the residual masses $M$ is a matter of convenience $Q = M^* - M$. The components $P_i$ are determined by the fragment emission angles under the assumption of conservation a projectile momentum per nucleon. In such an approach the contribution of the decays ${}^{9}$B $\to$ ${}^{8}$Be$p$ $\to$ 2$\alpha$$p$ in relativistic dissociation of the isotopes ${}^{10}$B and ${}^{10,11}$C is revealed  and, then, an indication to the resonance around 4 MeV in the channel ${}^{10}$C $\to$ ${}^{9}$B$p$ found \cite{2}. 

This experience gave confidence to search in the relativistic dissociation ${}^{12}$C $\to$ 3$\alpha$ for the Hoyle state (HS) or the second ${}^{12}$C exited state exceeding just 378 keV the 3$\alpha$-threshold \cite{3,4,5}. The HS studies are reviewed in \cite{6}. This search is inspired by the concept of the $\alpha$-particle Bose-Einstein condensate whose status is discussed most recently in \cite{6}. In the 90s, the ${}^{8}$Be contribution was determined by smallest $\alpha$-pair opening angles in ``white'' stars (coherent dissociation not accompanied by any target fragment or produced meson) ${}^{12}$C $\to$ 3$\alpha$ \cite{8} and ${}^{16}$O $\to$ 4$\alpha$ \cite{9} at energy of 3.65 $A$ GeV. Nowadays, the persisted NTE plates and the relevant data files \cite{8,9} are reused to obtain distributions over the $\alpha$-triple invariant mass $Q_{3\alpha}$. Data on the 72 (G.M. Chernov’s group, Tashkent) and 114 ``white'' stars ${}^{12}$C $\to$ 3$\alpha$ (A.Sh. Gaitinov’s group, Alma-Ata) underwent reanalysis. Besides, additional data on 238 3$\alpha$ stars including 130 ``white'' ones are obtained recently in this exposure. Then, the NTE layers exposed to 420 $A$ MeV ${}^{12}$C nuclei allow one to verify the invariant mass approach \cite{5}. In the latter case the $\alpha$-particle emission angles are measured in the 86 found 3$\alpha$ events including the 36 ``white'' stars.

\begin{figure}
	\centerline{\includegraphics*[width=0.75\linewidth]{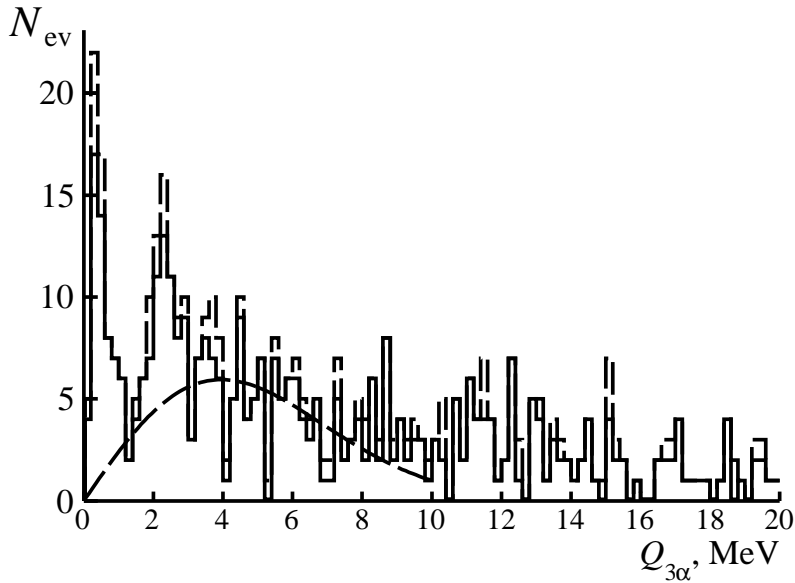}}
	\caption{Distribution over invariant mass $Q_{3\alpha}$ of all $\alpha$-triples in dissociation of ${}^{12}$C $\to$ 3$\alpha$ at 3.65 $A$ GeV (shaded) and 420 $A$ MeV (added by dashed line); line - approximation by the Rayleigh distribution with the parameter $\sigma_{Q_{3\alpha}}$ = (3.9 $\pm$ 0.4) MeV.}
\end{figure}

\begin{figure}
	\centerline{\includegraphics*[width=0.75\linewidth]{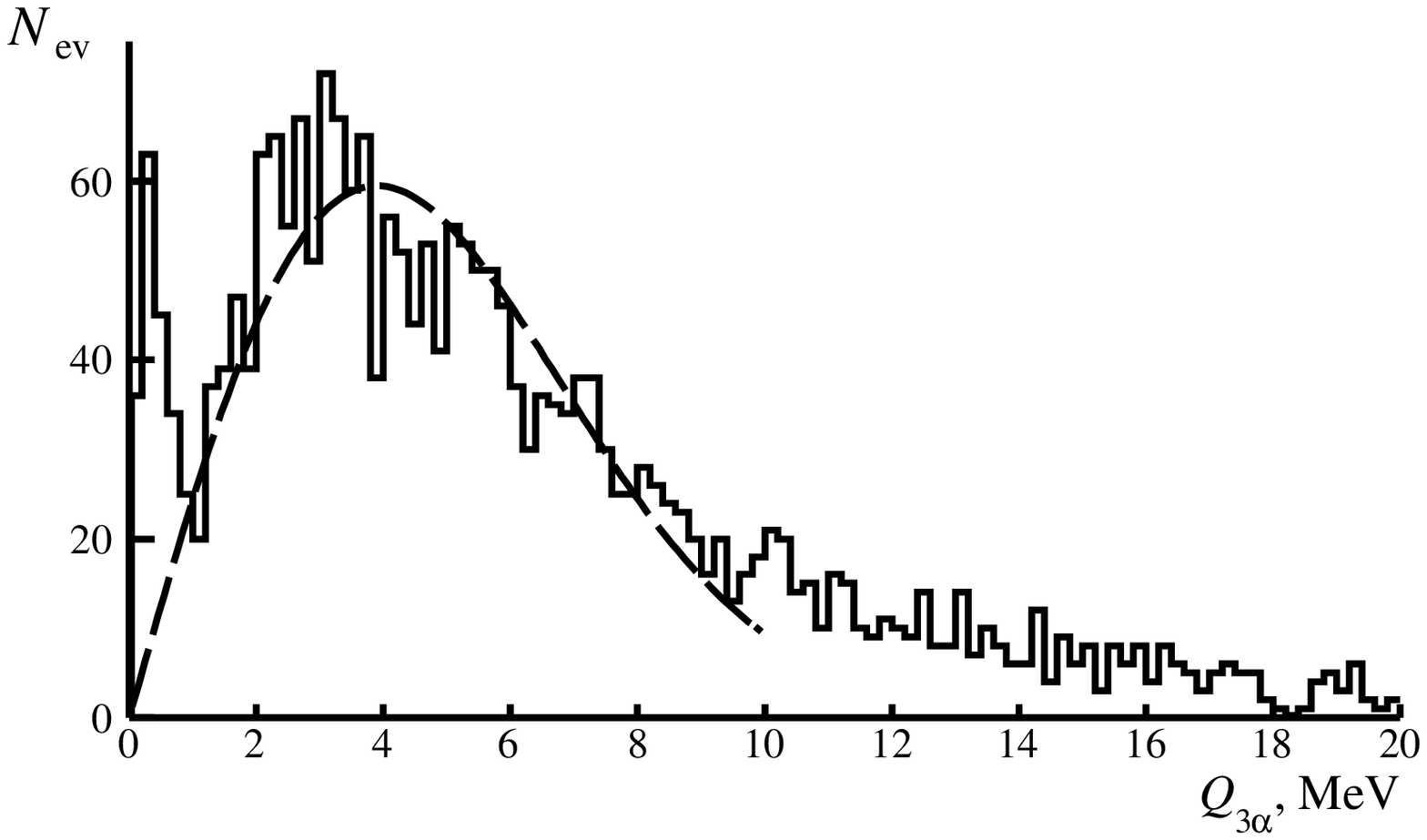}}
	\caption{Distribution over invariant mass $Q_{3\alpha}$ of all 3$\alpha$ combinations in 641 events of ``white'' stars ${}^{16}$O $\to$ 4$\alpha$; line - approximation by the Rayleigh distribution with the parameter $\sigma_{Q_{3\alpha}}$ = (3.8 $\pm$ 0.2) MeV.}
\end{figure}

The distribution $Q_{3\alpha}$ for all 510 stars is shown in Fig. 1. There is a peak in the region $Q_{3\alpha}$ $<$ 1 MeV for the 51 stars where HS signal is expected. For events at 3.65 $A$ GeV the mean value for the events contributed in the peak $\left\langle {Q_{3\alpha}} \right\rangle$ (RMS) is 397 $\pm$ 26 (166) keV, and at 420 $A$ MeV, respectively, 346 $\pm$ 28 (85) keV. According to the condition $Q_{3\alpha}$ $<$ 0.7 MeV 42 (of 424) events at 3.65 $A$ GeV can be attributed to HS and 420 $A$ MeV – 9 (of 86), including 5 ``white'' stars (of 36).  In sum, the contribution of the HS decays to the dissociation of ${}^{12}$C $\to$ 3$\alpha$ is (11 $\pm$ 3) \%. 

HS could emerge in the dissociation ${}^{16}$O $\to$ ${}^{12}$C$^*$ ($\to$ 3$\alpha$) + $\alpha$. Figure 2 shows the distribution $Q_{3\alpha}$ of all 3$\alpha$ combinations in the 641 ``white'' stars \cite{9}. While its main part limited $Q_{3\alpha}$ $<$ 10 MeV is described by the Rayleigh distribution there is the peak at $Q_{3\alpha}$ $<$ 700 keV. The condition $Q_{2\alpha}$ $<$ 200 keV meaning at least one ${}^{8}$Be per 4$\alpha$-event does not affect the statistics in this region. The main part contribution in the peak estimated at 8\% is excluded. The remaining 139 events have an average value $\left\langle {Q_{3\alpha}} \right\rangle$ = (349 $\pm$ 14) keV corresponding to HS and RMS 174 keV defined by the method resolution. In 9 of them more than one 3$\alpha$-combination meets the condition $Q_{3\alpha}$ $<$ 700 keV. In sum, the contribution of HS decays to coherent dissociation of ${}^{16}$O $\to$ 4$\alpha$ is (22 $\pm$ 2) \%. Besides, HS can arise the $\alpha$-decay product of the ${}^{16}$O 0$^{+}_{6}$ state \cite{7} (in an analogy with the HS decay ${}^{8}$Be + $\alpha$). The condition $Q_{4\alpha}$ $<$ 1 MeV allocates 9 events satisfying  with a mean value $\left\langle {Q_{4\alpha}} \right\rangle$ = (624 $\pm$ 84) keV at RMS 252 keV. Then, the estimate of the 0$^+_6$ contribution is (7 $\pm$ 2)\%.

Thus, HS identified in the relativistic ${}^{12}$C dissociation is manifested in the ${}^{16}$O case. These observations indicate that it is not reduced to the unusual ${}^{12}$C excitation and, like ${}^{8}$Be, is a more universal object of nuclear molecular nature. The closest confirmation of this assumption would be the HS observation in relativistic fragmentation ${}^{14}$N $\to$ 3$\alpha$. The analysis of the NTE layers exposed to relativistic ${}^{14}$N nuclei \cite{10} is resumed in the HS context. 

%
%

\end{document}